\documentstyle [12pt]{article}
\newcommand{\be}{\begin{equation}}
\newcommand{\ee}{\end{equation}}
\begin{document}
\pagestyle{empty}
%
%
%
\addtolength{\baselineskip}{0.20\baselineskip}
\hfill  IFUP-TH 11/94

\hfill   February 1994

\hfill  hep-lat/9403016

\begin{center}

\vspace{36pt}
{\large \bf Field--strength correlators in SU(2) gauge theory}

\end{center}

\vspace{36pt}

\begin{center}

{\sl L. Del\/ Debbio}, {\sl A. Di\/ Giacomo\/}
\vspace{12pt}

Dipartimento di Fisica dell'Universit\`a and I.N.F.N. \\
Piazza Torricelli, 2 \\
I-56100 Pisa, Italy \\
\vspace{16pt}

{\sl Yu. A. Simonov}
\vspace{12pt}

I.T.E.P. \\
B.Cheremushkinskaya ulitsa 25 \\
RU-117 259 Moskva, Russia\\

\end{center}

\vspace{36pt}

\begin{center}
{\bf Abstract}
\end{center}
\vspace{10pt}

We measure field-strength and their correlators in presence of a
static $q \bar q$ pair by numerical simulations. We give an
interpretation of these data in terms of quadratic and quartic
cumulants.

\vfill
\newpage
\clearpage
\pagenumbering{arabic}
\pagestyle{plain}

\section{Introduction}

A study of the field--strength distribution between static quark and
antiquark yields a detailed picture of the string formation and helps to
understand the mechanism of confinement. In particular, comparing the field
distributions with those of the dual Meissner model one can clarify the
viability of the popular confinement mechanism.

In non-abelian theory there are two ways of measuring the field--strength
distribution: using connected $\rho^c$ or disconnected $\rho^{disc}$
plaquette averages around the Wilson loop [1]. While both reduce to the same
quantity in the Abelian case, in the non-abelian case two measurements yield
independent information.
Recently [1] $\rho^c$ and $\rho^{disc}$ have been  measured for some
components of the field $F_{\mu\nu}$ using the cooling method [2]. While the
signal for $\rho^{disc}$ was too small, the distribution of $E_{11}$ off the
string axis ("the string profile") and along the  string from  $\rho^c$ was
found with good accuracy.
Another important quantity measured in [1] was a double plaquette correlator
$\gamma^c(x,x')$, yielding additional information on correlation of field
strength at two different points $x,x'$ of the string.

The purpose of the present paper is: i) to extend the Monte--Carlo (MC)
measurements to all orientations of the plaquettes in $\rho^c$ and
$\gamma^c(x,x')$, thus yielding the most complete information on the field
distribution in the $q\bar{q}$ string; ii) to give interpretation of the
results in terms of simpler quantities -- field strength correlators like
$\langle F_{\mu\nu}(x)F_{\rho\sigma}(y) \rangle$ (cumulants).

Using cluster expansion and non-abelian Stokes theorem [3], one can express
$\rho^c$ and $\gamma^c$ in terms of cumulants and keep the lowest ones
(physical arguments  in favour of dominance of lowest cumulants are given in
the review paper [4]). The latter have been recently measured on the lattice
[5], yielding a rather small correlation length $T_g=0.2~\mbox{fm}$.
Now, using the cumulants, one unambiguously predicts $\rho^c$ and can
compare it with MC measurements. This comparison with data from [1] was done
in [6] and a good agreement of measured and calculated string profile was
found. In particular it was clarified how a small correlation length
$T_g=0.2~\mbox{fm}$ yields a bigger radius of the string --
$0.5~\mbox{fm}$.

After presenting new measurements in sections 2 and 3, we
make in section 4 a  more precise and  extended calculation of $\rho^c$ in
terms  of the bilocal cumulant and compare it with  MC data. In particular
we predict vanishing of some $\rho^c(F_{\mu\nu})$ on symmetry grounds and
observe it explicitly in data.
We perform the same analysis for $\gamma^c$ and predict vanishing of
most combinations of plaquette orientations.
The dominating structure  -- $\gamma^c(E_{\Vert}, E_{\Vert})$ -- is
expressed in terms of the quartic cumulant, which has never been measured on
the lattice; the data for $\gamma^c$ allow to make some  estimates of it.

A short summary of results is given in conclusion.

\section{MC study of the field strength tensor.}

Using MC technique, we study the spatial distribution of the components of
the field strength tensor in presence of a $q\bar{q}$ pair. This generalizes
to all the components of the field the results already obtained in
[1]. Following~[1], we define:
\be
\rho_{\mu\nu}^c = \frac{\langle
tr(WLP_{\mu\nu}(x_{\Vert},x_{\bot})L^+) \rangle}{\langle Tr
W \rangle}-1
\ee
 where $W$ is a Wilson loop, $L$ is a Schwinger line and
$P_{\mu\nu}$ is the part of the plaquette proportional to the  $\sigma $--
matrices, oriented in order to give the desired component  of the field. The
coordinates $x_{\Vert}, x_{\bot}$ measure  resp. the distance from the
edge of the Wilson loop and from the plane defined by the loop, as
shown in Fig.~1.
In the naive continuum limit $a \to 0$,
\be
\rho^c_{\mu\nu}\simeq a^2 \langle F_{\mu\nu} \rangle_{q\bar{q}}
\ee
We have used a $16^4$ lattice, taking a $8\times 8$ Wilson loop and  $\beta
= 2.50$, which is inside the scaling window for the fields. Moving the
plaquette in and outside the plane defined by the Wilson loop, we obtain a
map of the spatial structure of the field as a function of $x_{\Vert}$ and
$x_{\bot}$. Using a controlled cooling technique (see [1,2] and references
therein), we eliminate the short-range fluctuations. The long--range
non--perturbative effects survive longer to the cooling procedure, showing a
plateau of 10--14 cooling steps, while the error becomes smaller. A similar
behaviour has been observed for the string tension. The cooling technique
allows us to disentangle the signal from the quantum noise with a relatively
small statistics. The general patterns of the field configurations are
briefly resumed in the following figures.Figure~2 represent a detailed map of
the spatial
behaviour of the longitudinal component of the chromoelectric field.

\begin{itemize}

\item Varying $x_{\Vert}$ at fixed $x_{\bot}$, we investigate the
structure of the
fields in the direction of the axis joining the $q\bar{q}$ pair. In
Fig.~3, we show $E_{\Vert}$ and $B_{\bot}$ as functions of $x_{\Vert}$
for $x_{\bot} = 0$,
i.e. on the $q\bar{q}$ axis. We find that the electric field remains
constant and magnetic transverse field vanishes, as expected on symmetry
grounds.

\item Varying $x_{\bot}$ at fixed $x_{\Vert}$, we study the transverse
shape of the fields.
Fig.~4 illustrates the behaviour of the $E_{\Vert}$ component vs.
$x_{\bot}$, for different values of $x_{\Vert}$: the field remains
constant with respect to $x_{\Vert}$ also outside the plane defined
by the Wilson loop, as long as we remain inside the string (i.e. for
$x_{\Vert}=3,5$). A detailed study of the transverse shape will be
given below.

\end{itemize}

We find that the parallel electric field is squeezed in flux tubes, as
already  found in [1]. The results in [1] were consistent with a
gaussian behaviour  of the flux tube profile inside the string. In
order to estimate the size  of our tubes and to check the consistency
of the result with previous  measurements, we have again performed a
fit of the transverse shape inside the string (for $x_{\Vert}=3$) with
the function
\be
E_{\Vert}=\exp(\kappa - \mu^2x^2_{\bot})
\ee
finding $\mu=0.30\pm0.01$, with $\chi^2 / \mbox{d.o.f.} = 0.993$. This
indicates that the
flux tubes have a transverse size of the order of 3 lattice spacings at
$\beta =2.50$, which corresponds to a physical value
$\mu^{phys}/\Lambda_{latt}=85 \pm 4$, in agreement with~[1].
In what follows, motivated by the measured form of the field strength
correlators~[5] and by the analysis in terms of cumulants~[3], we will
find that data are equally consistent with the form eq.~(18), which is
also suggested by the mechanism of confinement via dual superconductivity.

\section{Field strength correlators}

In the last years, a systematic study on non--perturbative effects in QCD
in terms  of the gluon field strength correlators has been developed (see
Ref. [3] and references therein) and  the  behaviour of these correlators in
the vacuum has been investigated by lattice simulations [5]. As pointed out
by the authors of [3], studying  the field correlators in presence of
$q\bar{q}$ pair, could provide further informations for describing color
confinement. Therefore we measure the operator:
\be
\gamma^c
=\frac{\langle Tr\{WSV_{PP'}S^+\}\rangle}{\langle Tr W
\rangle}-\langle TrV_{PP'} \rangle \ee
where
\be
V_{PP'}=P_{\mu\nu}LP'_{\rho\sigma}L^+-\frac{1}{2}P_{\mu\nu}TrP'_{\rho\sigma}
\ee
where $W$ is a Wilson loop, $S$ is a Schwinger line connecting the Wilson
loop to the $V_{PP'}$ operator, $P$ and $P'$ are two plaquettes, located at
$x$ and $x'$ respectively, and $L$ is a Schwinger line connecting them.

In the naive continuum limit, we have
\be
\gamma^c=a^2\langle F' \rangle +a^4[\langle FF' \rangle_{qq}- \langle
FF' \rangle_{0}]
\ee
where $F$ and $F'$ are respectively the field components at $x$ and $x'$.

Varying the orientations of the two plaquettes, we obtain the different
components of the correlators. The measurements have been done on a $12^4$
lattice at $\beta = 2.50$ using $6\times 6$ Wilson loop. Again, we used
controlled cooling to reduce the fluctuations. We have measured $\gamma^c$
with the following two types of orientations of $F$ and $F'$.

\noindent
(i) In the first case:

$\bullet~~P$  is held fixed on the $q\bar{q}$ axis at 1 lattice spacing
from the  border of the Wilson loop, while its orientation is varied in the
6 possible directions;

$\bullet~~P'$ is moved in- and outside  the plane of the Wilson loop, its
orientation is kept fixed in the $E_{\Vert}$ direction;

$\bullet~~x_{\Vert}$ and $x_{\bot}$ identify the position of $P'$ with respect
to
$P$.

\noindent
(ii) In the second type of measurements:

$\bullet$ both the position and the orientation of $P$ are kept fixed; the
plaquette is in the same position as before and its orientation corresponds
to the $E_{\Vert}$ component;

$\bullet~~P'$ is moved as before and its orientation is changed.

We finally define the irreducible correlator $\bar{\gamma}^c$ as follows
\be
\bar{\gamma}^c \equiv\gamma^c(x,x') - \rho^c(x') \approx
a^4[\langle FF'\rangle_{q\bar{q}}-\langle FF'\rangle_0]
\ee

\noindent
{}From (7) it is clear, that $\bar{\gamma}^c$ contains only  double plaquette
correlations.

\noindent
Most of the data for $\gamma^c$ and $\bar{\gamma}^c$ are
compatible with zero net effect, within two standard deviations. In
Table~1, we report  the  data
for  $\bar{\gamma}^c$ in case when  both plaquettes $P$ and $P'$ are kept
fixed in the  $E_{\Vert}$ direction. In the next Section we compare the
Monte--Carlo measurements with the predictions from the cumulant (cluster)
expansion and will see that indeed all orientations except $E_{\Vert},
E_{\Vert}$ should give zero result due to simple symmetry arguments.

\section{Extracting bilocal and quartic field-strength correlators from the
 Monte--Carlo data}

For the contour $C$ shown in Fig. 1 we denote direction
along the $q\bar{q}$ axis $x_{\Vert} = x_1$,
while that of $x_{\bot} = x_2$, and the Euclidean temporal axis is $x_4$. All
the construction in Fig. 1 is taken at a fixed value of $x_3$.

Using the non-abelian Stokes theorem and the cluster expansion  [3] for
$\rho^c_{\mu\nu}$  in (1) one has (see [6] for details of derivation)
\be
\rho^c_{\mu\nu}(x_1,x_2,x_4)= a^2\int  d\sigma_{14}(x'_1,x'_4)
\Lambda_{\mu\nu}
\ee
where
\be
\Lambda_{\mu\nu} =\frac{1}{N_c}tr \langle E_1(x'_1,0,x'_4)\Phi
F_{\mu\nu}(x_1,x_2,x_4) \Phi^+ \rangle +...,
\ee
$\Phi$ is the parallel transporter (Schwinger line) from the point $(x'_1,0,
x'_4)$ to $(x_1,x_2,x_4)$, and dots imply contribution of
higher order cumulants, containing additional powers of $E_1$
[3].

We shall keep throughout this Section only the lowest cumulants
(containing lowest power of $E_1$) and   compare our prediction
with the MC data of previous sections. The  bilocal correlator
$\Lambda_{\mu\nu}$  can be expressed in terms of two independent
Lorentz scalar functions $D((x_{\mu}-x'_{\mu})^2)$,
$D_1((x_{\mu}-x'_{\mu})^2)$ (see [3] and the appendix 1 of the
last ref. in [3])
\be
\Lambda_{14}=D+D_1+(h^2_1+h^2_4)\frac{dD_1}{dh^2}
\ee
\be
\Lambda_{24}=(h_1h_2)\frac{dD_1}{dh^2}~~,~~~
\Lambda_{34}=(h_1h_3)\frac{dD_1}{dh^2}
\ee
\be
\Lambda_{23}\equiv 0,~~
\Lambda_{13}=h_3h_4\frac{dD_1}{dh^2}~;~~
\Lambda_{12}=h_2h_4\frac{dD_1}{dh^2}
\ee
Here $h_{\mu}=(x-x')_{\mu}$.

\noindent
Since all construction in Fig. 1 is at $x_3=x'_3=0$ we have $h_3\equiv 0$
and hence
\be
\rho^c_{23}=\rho^c_{34}=\rho^c_{13} \equiv 0
\ee
The only nonzero components are $\Lambda_{14}, \Lambda_{24}$ and
$\Lambda_{12}$. For the latter the contribution to $\rho^c$ can  be written
as
\be
\rho^c_{12}(x_1,x_2,x_4)= a^2\int^R_0 dx'_1\int^{\frac{T}{2}}_{-\frac{T}{2}}
dx'_4 (+x_2)(x_4-x'_4)\frac{dD_1(h^2)}{dh^2}
\ee
When $x_4=0$ ( and this is where measurements of $\rho^c_{12}$ have been
done), $\rho^c_{12}$ vanishes  because of antisymmetry of the integrand in
(14).

\noindent
Hence only $\rho^c_{14} $ and $\rho^c_{24}$ are nonzero, and only those have
been measured to be nonzero.

To make comparison with data more  quantitative,  let  us exploit recent MC
calculation of $D$ and $D_1$ [5], which imply  that both $D$ and $D_1$ are
of exponential form
\be
D_1(h^2)= D_1(0)\exp(-\mu_1 h); D(h^2) = D(0) \exp (-\mu h)
\ee
$$D_1(0) \approx \frac{1}{3} D(0);~~ \mu_1\approx \mu $$
Inserting this into (8), (14) we have
\be
\rho^c_{14}(x_1,x_2;0)= a^2\int^R_0
dx'_1\int^{\frac{T}{2}}_{-\frac{T}{2}} dx'_4
[D(0)+D_1(0)-\frac{(h^2_4+h^2_1)}{2h}D_1(0)]e^{-\mu h}
\ee
with
$$h_4=-x'_4~,~~h_1=x_1-x_1'~~,~~ h^2=h^2_4+h^2_1+x^2_2;$$

\noindent
For $\rho^c_{24}$ similarly one obtains
\be \rho^c_{24}(x_1,x_2;0)= -a^2\mu
x_2 \int^R_0 dx'_1\int^{\frac{T}{2}}_{-\frac{T}{2}} dx'_4
\frac{(x_1-x_1')}{2h}D_1(0)e^{-\mu h}
\ee
{}From (16) and (17) one can deduce, that \\
(i) $\rho^c_{24}$ should vanish for $x_2=0$\\
(ii) $\rho^c_{24}$ changes sign for  $x_1=\frac{R}{2}$, i.e. in the middle
of the string length.\\
(iii) $\rho^c_{24}$ is about $1/3$ of  $\rho^c_{14}$.

\noindent
All properties
(i)--(iii) are supported by the data.

\noindent
Finally, we can make a detailed comparison of our prediction for
$\rho^c_{14}$ in (16) with data. One
obtains a simple analytic result  for $\rho^c_{14}(x_2\equiv
x_{\bot})$ in case of a very long string.  The transverse shape
measured at the middle is given
by [6]

\be
\rho^c_{14}=\frac{2\pi a^2}{\mu^2}[D(0)(1+\mu x_2) -
D_1(0)\frac{1}{2} (\mu x_2)^2]e^{-\mu x_2}
\ee
As  shown in [6], this shape is in good agreement with the  previous
data, obtained in [5].
Here we calculate $\rho^c_{14}$ as a function of $x_1,x_2$ from (16)
keeping $D_1(0)=\frac{1}{3} D(0)$.
We then fit the data for $x_{\Vert}=3$ to evaluate $\mu$ and $a^2
D(0)$.
We find:
\[
\mu \approx 0.19~\mbox{fm},~~~~ a^2 D(0) \approx 3.92 \times
10^7
\]
with a $\chi^2 / \mbox{d.o.f} = 0.17$.

\noindent
The value  of $\mu$ is in good agreement with [5], while we find that
$a^2~D(0)$ is one order of magnitude smaller than in the previous
measurements. We recall that our data here are obtained for $SU(2)$,
while in~[5] the gauge group was $SU(3)$. This should account for the
one order of magnitude between the two results.

\noindent
These results allow to predict all curves for other values of
$x_{\Vert}$ and $x_{\bot}$: the agreement with the numerical results
is very satisfactory as can be seen from Fig.~4.

We turn now to the double correlator
$\bar{\gamma}^c_{\mu\nu}$, Eq. (7).
We again use the non-abelian Stokes theorem
and the cluster expansion, to represent
$\bar{\gamma}^c$ as
\be
\bar{\gamma}^{c}_{\mu\nu,\mu'\nu'}(x,x')=\frac{a^4}{4!} \int dy_1
dy_4du_1du_4\{\ll E_1(y)\Phi E_1(u)\Phi F_{\mu\nu}(x)\times
\ee
$$\Phi F_{\mu'\nu'}(x')\Phi\gg+ \: perm \}$$
$$+a^4\frac{(-i)}{3!}\int dy_1dy_4\{\ll E_1(y)\Phi
 F_{\mu\nu}(x)\Phi
F_{\mu'\nu'}(x') \Phi\gg+ \: perm \}
$$
where the sum is over permutations of the order in which $E_1$ and
$F_{\mu\nu}$ appear under the sign of the cumulant; the latter is denoted by
double angular brackets and implies that
vacuum insertions are subtracted from the
averages of the field strengths.

\noindent
In our MC calculations partly reported in
Table 2, the orientation of the plaquette $P$ or $P'$ was kept along
the plane 14, and we should consider accordingly in (19) always $E_1(x)$ or
$E_1(x')$ inside the cumulants.

\noindent
Now the symmetry requirements impose severe
conditions on the nonzero values of
$\bar{\gamma}_{\mu\nu,\mu'\nu'}^{c}$.
Since both $P$ and $P'$  are chosen in
the middle of the $x_4$ interval for the
Wilson loop $[-\frac{T}{2},\frac{T}{2}]$, one can use the
symmetry with respect to the change
$x_4\to -x_4$. In this way  one can show
that all odd--power averages of the type
$\ll E_1(u)\Phi E_1(y)\Phi
...E_1(P')\Phi\gg$ should vanish, since
they are odd with respect to $x_4\to
-x_4$.

\noindent
This property holds if one inserts in this odd--power cumulant
additionally several magnetic field operators $B_i, B_k...$

\noindent
Similarly, the correlator containing $B_i(P)$ should depend on , e.g.  \be
\ll E_1(u)\Phi B_i(x)\Phi
E_1(x')\Phi\gg \sim e_{ikl}h_kh_l
\ee
and therefore should  vanish whenever $h_k$ or $h_l$ are zero for both
intervals, with $h_{\mu}=(u-x)_{\mu}, ~~h'_{\mu}= (u-x')_{\mu}$.
In this way one proves  that (20) vanishes for $i=1$ or 2 identically,
since $h_3=h'_3\equiv 0$. For $i=3$ the combination (20) should vanish
when $h_2=h'_2=0$. Using those criteria we keep in our Table 1 only
results for  $\gamma_{14,14}$, where the signal is largest. The
contribution to $\gamma_{14,14}$ comes only  from the quartic cumulants
in (19), since as we discussed  above, the triple correlator
$\ll E,\Phi E_1\Phi E_1\Phi\gg$ vanishes identically .

One can conclude from the Table 1, that the quartic cumulant  sharply
decreases for large $x_{\bot}$, and the transverse shape of the string,
using $\bar{\gamma}_c$, is similar to that deduced from $\rho_c$, the
thickness of the string being of the order of 3 lattice spacings.

All the correlations $\gamma_{\mu\nu,\mu'\nu'}$, which should vanish
on symmetry grounds, have been found to be zero within two
standard deviations. In these cases, a higher statistics should be
necessary for a clear answer.

\section{Conclusions}
One should stress three different aspects of the results reported above.
First of all, we have presented the most detailed measurements of the
connected correlators made so far. There is  an agreement between the field
distributions of this paper and those reported in [1], where only
correlators of $E_{11}$ were measured to be nonzero. Here with better
statics all components of magnetic  field $\vec{B}$ and electric field
$\vec{E}$ are given for all possible configurations of one probing
plaquette (for $\rho_c$) and two probing plaquettes (for $\bar{\gamma}^c$).
Symmetry requirements impose severe restrictions on $\rho_c$ and
$\bar{\gamma}^c$ (independently on cluster expansion) and predict zero
results in most cases, expect for $E_{11}(\rho^c_{14}), E_2(\rho^c_{24})$
and for $\bar{\gamma}^c(E_{11}, E_{11})$. Results are compatible with the
predictions, and hence statistical errors seem to be reliable.

Secondly, our results serve to
check the validity and usefulness of the cluster expansion for the signals
like $\rho^c$, $\bar{\gamma}^c$. This expansion allows to express $\rho^c$
and $\bar{\gamma}^c$ in terms of a simpler (and more fundamental) bilocal
correlator $G_2\equiv \langle F\Phi F\Phi \rangle$, which has been
measured earlier [5] and
thus yields a clear prediction for $\rho^c$ and $\bar{\gamma}^c$. Comparison
with measured results made earlier in [6] and here in more detail in Fig.
3,4,  shows a good agreement and supports the  fundamentality of the bilocal
correlator, which is known to define the nonperturbative dynamics of
confinement [3].

In particular the asymptotics of the string profile at large $x_{\bot}$ is
shown to be exponential, see eq. (18), just as the asymptotics of $G_2$,
measured in [5].  This in contrast to the behaviour inside the string, where
the Gaussian--like  flattening is observed before [1] and also in this
paper. This effect is connected to the flattening of $G_2(x)$ at small $x$,
necessary for its  regularity at $x=0$ (Note that the latter property of
 $G_2$ was not taken into account in (15)) and  also because of the smearing
effect due to integration in (16), yielding polynomial factors in  (18).

Of special importance is the first estimate of  quartic cumulant
$G_4\equiv \langle (F\Phi)^4 \rangle$ through $\bar{\gamma}^c$ in (19).

To obtain a more quantitative measure of $G_4$, we represent the coordinate
dependence of $G_4$ as an exponent similarly to $G_2$ (15) with the same
$\mu$. In this case the dimensionless ratio is
$$\frac{G_4(0)}{(G_2(0))^2}=\frac{\bar{\gamma}^c(0)}{\rho^c(0))^2}\approx
2-3$$

\noindent
This kind of estimate would result from the instanton--type vacuum with
instantons of small size $\rho, \rho\leq 0.3 fm$ and typical
density of one instanton per $fm^4$~[7].

Finally, let us compare the transverse shape of the string (the string
profile) measured in Table 1 and analytically written in (18) with that of
the dual Meissner effect.
As is known from the theory of the type-II superconductors [8,
Eq.(48.11)] the asymptotics of the magnetic field distribution off the
vortex line is exponential, $B(r)\sim exp (-r/\delta)$, where $\delta$ is
the so-called
London's
penetration length. One can see that the dual Meissner effect is capable of
reproducing such fine details of the confinement picture, as the field
distribution around the string.

This conclusion agrees with a recent study made in a completely different
approach in  [9].

\vspace{2cm}

\underline{\bf Acknowledgements}

This work has been started, when one of the authors (Yu.S.)  was
visiting Department of Physics of Pisa University. It is a pleasure
for him to thank the Department for a kind hospitality, and to acknowledge a
financial support of the Russian Fund for Fundamental Research.
LDD wants to thank the Italian ALEPH group at CERN for the warm
hospitality during the final part of this work.

 \clearpage
 {\bf Figure Caption}
 \begin{itemize}
  \item Figure~1: The operator $\rho^c$.

\item Figure~2: The spatial structure of the parallel chromoelectric
field $E_{\Vert}$.

\item Figure~3: Different components of the field--strength tensor
($E_{\Vert}$ and $B_{\Vert}$, circles and diamonds respectively) vs.
$x_{\Vert}$ at fixed $x_{\perp}=0$.

\item Figure~4: $E_{\Vert}$ vs. $x_{\perp}$, for
$x_{\Vert}=1,3,5,7$ (triangles up, filled circles, diamonds and
triangles down resp.).  The dashed and solid lines are the results of
the fits performed using resp. formula~(18) and (3).
\end{itemize}
\clearpage
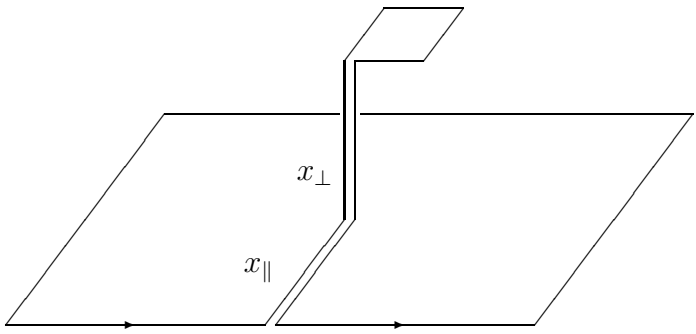
\begin{figure}[h]
\vspace{5cm}
\begin{picture}(250,140)
\put (50,0) {\line(1,0){98}}
\put (99,0) {\huge \vector(1,0){0}}
\put (152,0) {\line(1,0){98}}
\put (201,0) {\huge \vector(1,0){0}}
\put (148,0) {\line(3,4) {30}}
\put (152,0) {\line(3,4) {30}}
\put (178,40) {\line(0,1) {60}}
\put (182,40) {\line(0,1) {60}}
\put (178,100) {\line(3,4) {15}}
\put (193,120) {\line(1,0) {30}}
\put (208,100) {\line(3,4) {15}}
\put (182,100) {\line(1,0) {26}}
\put (250,0) {\line(3,4){60}}
\put (160,55) {$x_{\perp}$}
\put (140,20) {$x_{\parallel}$}
\put (110,80) {\line(1,0){66}}
\put (184,80) {\line(1,0){126}}
\put (50,0) {\line(3,4){60}}
\end{picture}
\vspace{2cm}
\caption{The operator $\rho^{c}$.}
\end{figure}
\clearpage
\begin{figure}[h]
\includegraphics{new_fig2.ps}
\vspace{17cm}
\caption{Spatial distribution of $E_{\parallel}$}
\end{figure}
\clearpage
\begin{figure}[t]
\includegraphics{fig3.ps}
\vspace{14cm}
\caption{$E_{\parallel}, B_{\parallel}$ vs. $x_{\parallel}$, for
$x_{\perp }=0$.}
\end{figure}
\clearpage
\begin{figure}
\includegraphics{terza_fig4.ps}
\vspace{14cm}
\caption{$E_{\parallel}$ vs. $x_{\perp}$.}
\end{figure}
\clearpage
\begin{tabular}{||c|c||} \hline
$x_{\perp}$ & $E_{\parallel}$  \\                                    \hline

 0 & -.032893 $\pm$ 0.008056 \\
 1 & -.029240 $\pm$ 0.006531 \\
 2 & -.020212 $\pm$ 0.005298 \\
 3 & -.010907 $\pm$ 0.004644 \\
 4 & -.005311 $\pm$ 0.004136 \\
 5 & -.001269 $\pm$ 0.004222 \\
 6 & 0.000397 $\pm$ 0.003870 \\
\hline
\multicolumn{2}{l}{Table 1: $\bar\gamma^{c}$ vs. $x_{\perp}$,
$x_{\parallel}=1$.}
\end{tabular}

\end{document}